\documentclass[sigconf]{acmart}

\setcopyright{acmcopyright}
\copyrightyear{2023}
\acmYear{2023}

\acmConference[WWW '23]{Proceedings of the ACM Web Conference 2023}{May 1--5, 2023}{Austin, TX, USA}
\acmBooktitle{Proceedings of the ACM Web Conference 2023 (WWW '23), May 1--5, 2023, Austin, TX, USA}
\acmPrice{15.00}
\acmDOI{10.1145/3543507.3583277}
\acmISBN{978-1-4503-9416-1/23/04}

\usepackage{enumitem}
\usepackage{multirow}
\usepackage{subfigure}
\usepackage{bbding}
\usepackage{balance}
\usepackage{bm}
\usepackage{color}

\usepackage[linesnumbered,ruled,vlined]{algorithm2e}
\SetKwInput{KwInput}{Input}                
\SetKwInput{KwOutput}{Output}              

\begin{document}

\title{ConsRec: Learning Consensus Behind Interactions for Group Recommendation}

\settopmatter{authorsperrow=4}

\author{Xixi Wu}
  \email{21210240043@m.fudan.edu.cn}
  \affiliation{%
    \institution{Shanghai Key Laboratory of Data Science, School of Computer Science, Fudan University}
    \city{Shanghai}
    \country{China}
  }

\author{Yun Xiong}
\authornote{Corresponding author}
 \email{yunx@fudan.edu.cn}
  \affiliation{%
     \institution{Shanghai Key Laboratory of Data Science, School of Computer Science, Fudan University}
     \city{Shanghai}
    \country{China}
  }
 
\author{Yao Zhang}
  \email{yaozhang@fudan.edu.cn}
  \affiliation{%
     \institution{Shanghai Key Laboratory of Data Science, School of Computer Science, Fudan University}
     \city{Shanghai}
      \country{China}
  }
 
\author{Yizhu Jiao}
  \email{yizhuj2@illinois.edu}
  \affiliation{%
    \institution{University of Illinois at Urbana-Champaign}
    \state{IL}
    \country{USA}
  }

   \author{Jiawei Zhang}
  \email{jiawei@ifmlab.org}
  \affiliation{%
     \institution{IFM Lab, Department of Computer Science, University of California, Davis}
     \state{CA}
     \country{USA}
  }

 \author{Yangyong Zhu}
  \email{yyzhu@fudan.edu.cn}
  \affiliation{%
     \institution{Shanghai Key Laboratory of Data Science, School of Computer Science, Fudan University}
     \city{Shanghai}
     \country{China}
  }
 
\author{Philip S. Yu}
  \email{psyu@uic.edu}
  \affiliation{%
    \institution{University of Illinois at Chicago}
    \state{IL}
    \country{USA}
}

\renewcommand{\shortauthors}{Xixi, et al.}

\begin{abstract}
Since group activities have become very common in daily life, there is an urgent demand for generating recommendations for a group of users, referred to as \textit{group recommendation} task. Existing group recommendation methods usually infer groups' preferences via aggregating diverse members' interests. Actually, groups' ultimate choice involves compromises between members, and finally, an agreement can be reached. However, existing individual information aggregation lacks a holistic group-level consideration, failing to capture the consensus information. Besides, their specific aggregation strategies either suffer from high computational costs or become too coarse-grained to make precise predictions.

To solve the aforementioned limitations, in this paper, we focus on exploring consensus behind group behavior data. To comprehensively capture the group consensus, we innovatively design three distinct views which provide mutually complementary information to enable multi-view learning, including member-level aggregation, item-level tastes, and group-level inherent preferences. To integrate and balance the multi-view information, an adaptive fusion component is further proposed. As to member-level aggregation, different from existing linear or attentive strategies, we design a novel hypergraph neural network that allows for efficient hypergraph convolutional operations to generate expressive member-level aggregation. We evaluate our ConsRec on two real-world datasets and experimental results show that our model outperforms state-of-the-art methods. An extensive case study also verifies the effectiveness of consensus modeling.

\end{abstract}

\ccsdesc[500]{Information systems~Recommender systems}
\ccsdesc[500]{Computing methodologies~Neural networks}

\keywords{Group Recommendation, Graph Representation Learning, Data Mining}

\maketitle

\section{Introduction}

\begin{figure}[t]
    \includegraphics[width=8cm]{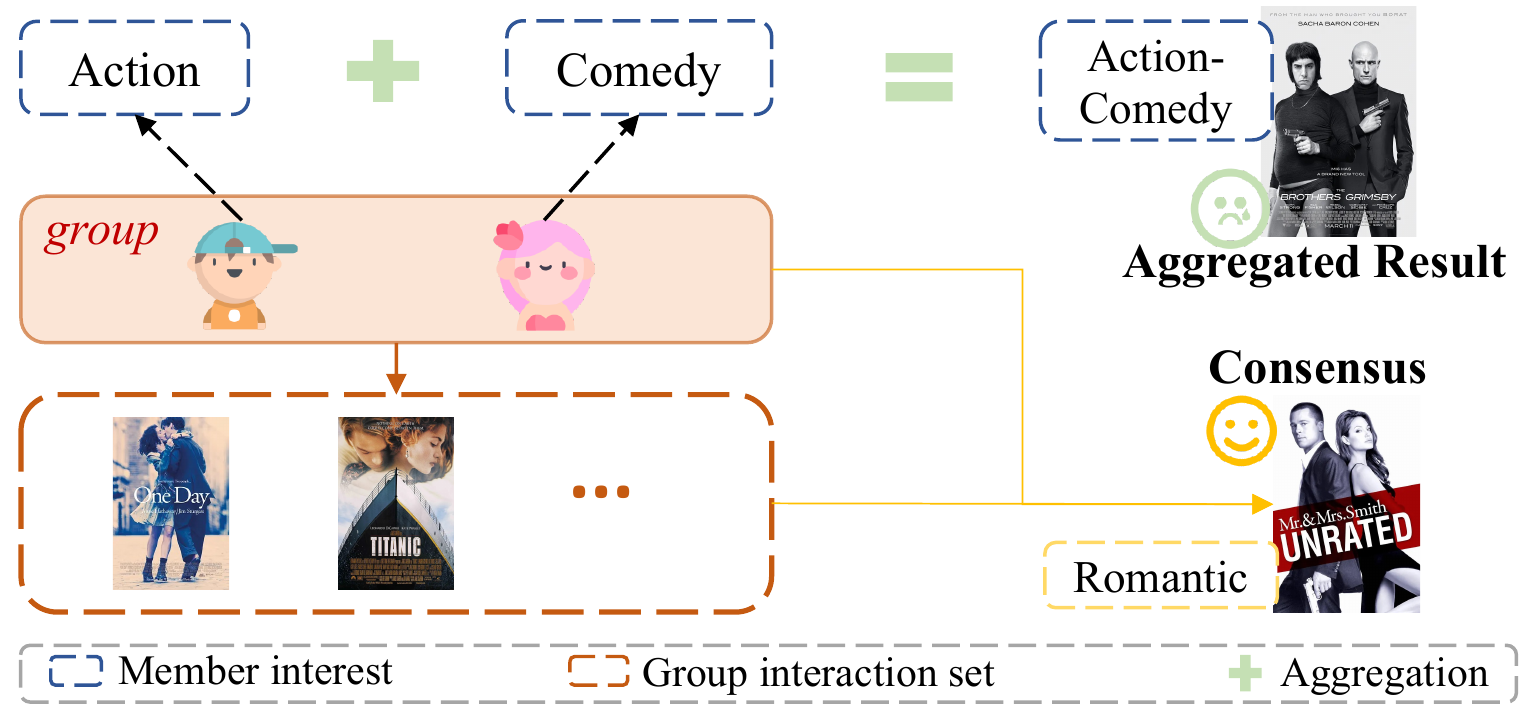}
    \caption{An illustrative example of the gap between aggregated result and group's consensus. Merely aggregating diverse members' interests lacks the holistic consideration of the group's overall taste, failing to capture the consensus.}
    \label{fig:example}
\end{figure}

Owing to the prevalence of social media \cite{SocialMedia1, SocialMedia2}, online group activities have become very common. Various existing online social media sites can provide group-related services for users. For example, travel enthusiasts can plan group trips on Mafengwo\footnote{https://www.mafengwo.cn}, and young people can also organize group parties on Meetup\footnote{https://www.meetup.com}. As traditional recommendation methods only target suggesting relevant items to individuals, there is an urgent demand in generating recommendations for a group of users, referred to as the \textit{group recommendation} task \cite{AGREE, SoAGREE}. Formally, group recommendation aims to reach an agreement among group members to provide contents that can satisfy most members \cite{HHGR}.

Previous works \cite{average, least_misery, MaxiSatisfaction, AGREE, SoAGREE, MoSAN, Voting, GAME, GroupIM, HyperGroup, HCR, HHGR, HyperCube, COM, PIT, UL_ALL} on group recommendation usually focus on aggregating diverse members' preferences to predict group interests. Traditional aggregation methods are based on pre-defined heuristic rules, including the average \cite{average}, the least misery \cite{least_misery}, and the maximum satisfaction \cite{MaxiSatisfaction}. There also exist several deep-learning based models \cite{AGREE, SoAGREE, MoSAN, Voting, GAME} that incorporate the attention mechanism to realize a learnable aggregation instead. Recently, CubeRec \cite{HyperCube} utilizes the geometric expressiveness of hypercubes \cite{box, box2} to aggregate multi-faceted member preferences, achieving state-of-the-art performance.

Despite their effectiveness, we argue that current group recommendation methods fall short in modeling two factors: (1) \textbf{Consensus}. An important fact has been neglected: groups' ultimate choice involves compromises between members, and finally, an agreement need to be reached \cite{HyperCube, Voting}. The consensus drives the group to make the final decision while existing individual information aggregation lacks a holistic group-level consideration, failing to capture the consensus. Taking Figure \ref{fig:example} as an example, the group consists of a couple where the man prefers action films while the lady enjoys comedies. When they plan to watch a movie together, the consensus may be a romantic movie rather than the aggregated action-comedy. Such situations are quite common in real scenes but are largely overlooked by the aforementioned methods. (2) \textbf{Member-level Aggregation}. A typical strategy for group interests aggregation is the attentive mechanism \cite{AGREE, SoAGREE, GAME, Voting, MoSAN, HCR} that calculates different attention weights for different candidate items. Considering the huge amount of candidate products in real-world scenarios, such methods are hardly applicable. Besides, this approach tends to favor members with frequent interactions since they are likely to obtain higher attention weights \cite{AttentionBias}. Though CubeRec \cite{HyperCube} resorts to hypercubes for a better solution, the obtained hypercube may be too large due to diversity of member interests to make precise predictions. Therefore, an appropriate aggregation strategy that realizes both efficiency and effectiveness is also demanded.

In this paper, we focus on exploring group consensus behind group-item and user-item interactions, so as to improve the performance and meaningfulness of recommendation. We propose a new model, \textbf{ConsRec}, to solve the foregoing limitations correspondingly: (1) \textbf{Group Consensus Modeling}. To comprehensively capture the group consensus, we innovatively design three distinct views which provide complementary information. Besides the member-level view that provides fine-grained explanation for group consensus, we argue that sometimes the consensus of groups can be captured from the characteristics of interacted items as well as their inherent attributes. For example, a group formed by a family is more inclined to watch family-style comedies. Therefore, we suggest the novel item-level and group-level views to obtain groups' item-level tastes and inherent interests, respectively. For each data view, we design a specific graph structure to encode the behavior data and utilize graph representation learning techniques to generate representations for groups. To integrate and balance multi-view information, an adaptive fusion component is further proposed to synthesize the final group consensus. During optimization, recommendation predictions can supervise the learning process to adaptively adjust the contributions of different views to extract the most discriminative consensus information. (2) \textbf{Effective Member-level Aggregation}. We propose a novel hypergraph learning architecture to obtain member-level aggregation. Compared with existing attentive aggregation, our schema wins in efficiency as obtaining aggregation via hypergraph convolutional operations, fairness as designing meaningful side information for balancing the favor of some members, and expressiveness as incorporating high-order collaborative information. Extensive experiments on two public datasets show the effectiveness as well as efficiency of our proposal.

To summarize, the contributions of our work are as follows: 
\begin{itemize}[leftmargin=*, topsep=2pt]
   \item We study the group recommendation task and reveal the consensus behind groups' decision-making. To comprehensively capture the consensus, we design three novel and complementary views, including member-level aggregation, item-level tastes, and group-level inherent interests. 

   \item We propose a novel hypergraph neural network to obtain member-level aggregation. Compared with existing attentive aggregation, our method wins in efficiency, fairness, and expressiveness.

   \item Extensive experiments on two public datasets show the effectiveness as well as efficiency of our proposal.
  
\end{itemize}

\section{Related Work}

\subsection{Preference Aggregation}

Existing group recommendation techniques usually follow the aggregation strategy, which first learns members' preferences from user-item interactions, and then performs preferences aggregation to infer the overall interest of a group \cite{HyperCube}. Sometimes, group's ultimate choice is determined by group-level consensus. However, existing methods lack the holistic consideration of the group’s overall taste, failing to capture the consensus.

\subsubsection{Score Aggregation} 
The approaches of this category generate the scores of all members in a group for a candidate item and then aggregate individual scores to obtain the preference score of the group through some hand-crafted heuristics, such as the average \cite{average}, the least misery \cite{least_misery}, and the maximum satisfaction \cite{MaxiSatisfaction}. For example, the average \cite{average} strategy takes the average score across members as the final recommendation score. Though intuitive, these pre-defined aggregation rules are inflexible to reflect diverse member intentions \cite{HCR}.

\subsubsection{Neural Aggregation} For better performance, some researchers propose to aggregate different members' intentions in a learnable way \cite{COM, UL_ALL, AGREE, SoAGREE, Voting, MoSAN, GAME, HyperCube}. Technically, these methods firstly represent each member as an implicit embedding based on user interactions, and then calculate the group embedding by summing the embeddings of all members with the learned weights. For example, attentive neural networks are proposed in \cite{AGREE} to selectively aggregate user representations within a group, and \cite{MoSAN} further captures the fine-grained interactions between group members via a sub-attention network. Recently, CubeRec \cite{HyperCube} utilizes the geometric expressiveness of hypercubes to adaptively aggregate the multi-faceted user preferences, achieving better performance.

\subsection{Recent Directions}
For better representation, recent studies on group recommendation resort to more complex structures, such as hypergraphs \cite{HCR, HyperGroup, HHGR} and hypercubes \cite{HyperCube}. Besides, self-supervised learning techniques are also adopted to counteract the data-sparsity issue \cite{GroupIM, HHGR}.

\subsubsection{Hypergraph Learning} As a more general topological structure that preserves the tuple-wise relationship, hypergraph naturally fits group profiling. Therefore, recent works \cite{HyperGroup, HCR, HHGR} represent the group as a hyperedge and employ hypergraph neural networks to generate more expressive vectors. We point out that these methods mainly utilize hypergraph structures to propagate user-level collaborative signals, and exhaustively employ the attentive aggregation \cite{attention} for group-level prediction. In fact, expressive member-level aggregation can be obtained more efficiently from hypergraph networks, which is neglected by these methods.

\subsubsection{Self-supervised Learning} 

As self-supervised learning (SSL) has shown its effectiveness in general recommendation tasks \cite{SGL, Yu2021SelfSupervisedMH, Zhou2020S3RecSL}, attempts are also made to design SSL objectives to counteract the data sparsity issue for group recommendation task \cite{GroupIM, HHGR, HyperCube}. For example, GroupIM \cite{GroupIM} proposes a user-group mutual information maximization scheme to jointly learn informative user and group representations. 

\begin{figure*}[!t]
    \centering
    \includegraphics[width=17cm]{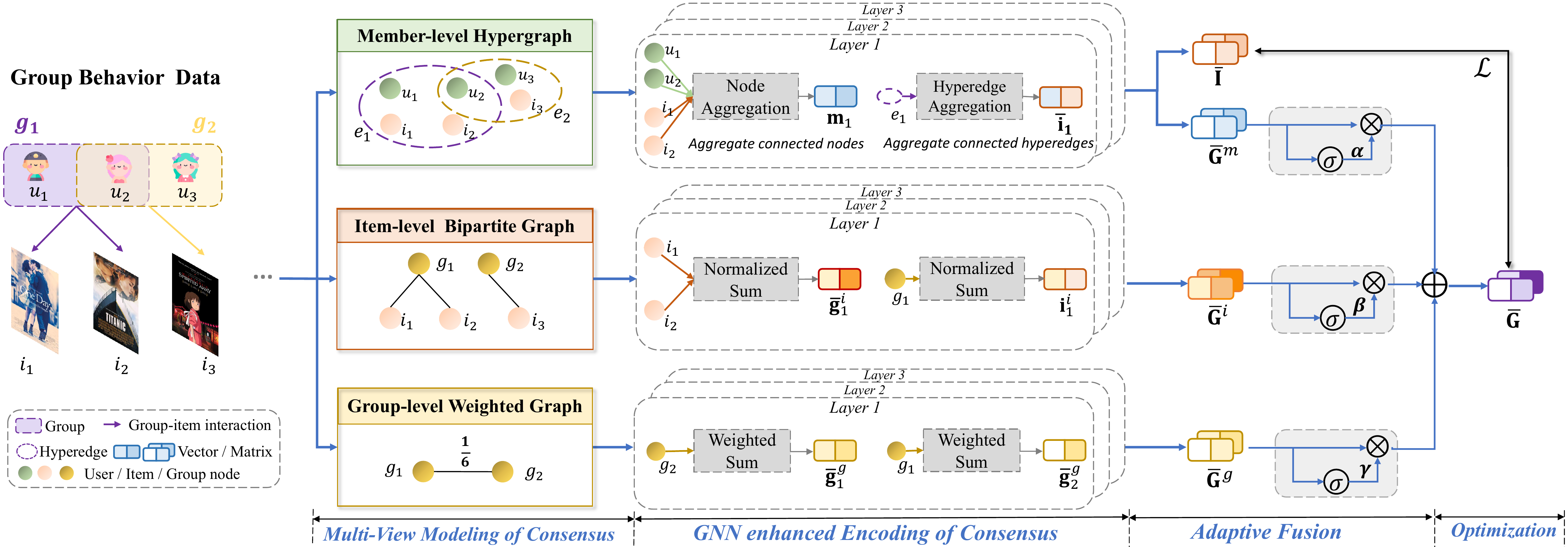}
    \caption{ConsRec Overview. We construct three distinct views for consensus modeling and adopt specific graph neural networks for representation learning. We further integrate these view-specific representations for group-item prediction.}
    \label{fig:overview}
\end{figure*}

\section{Preliminary}

In this section, we present the definition of group recommendation task and the concepts of hypergraph to facilitate comprehension. Formally, we use bold capital letters (e.g., $\mathbf{X}$) and bold lowercase letters (e.g., $\mathbf{x}$) to represent matrices and vectors, respectively. We employ non-bold letters (e.g., $x$) to denote scalars, and calligraphic letters (e.g., $\mathcal{X}$) to denote sets. Notations are summarized in Table \ref{tab:notation} in the Appendix.

\subsection{Task Definition}
\par Let $\mathcal{U}=\{ u_1, u_2, ..., u_M \}$, $\mathcal{I}=\{ i_1, i_2, ..., i_N \}$, and $\mathcal{G}=\{ g_1, g_2, ..., g_K \}$ be the sets of users, items, and groups, respectively, where $M$, $N$, and $K$ are the sizes of these three sets. There are two types of observed interactions among $\mathcal{U}$, $\mathcal{I}$, and $\mathcal{G}$, namely, group-item interactions and user-item interactions. We use $\mathbf{Y} \in \mathbb{R}^{K \times N}$ to denote the group-item interactions where the element $\mathbf{Y}(t,j)=1$ if group $g_t$ has interacted with item $i_j$ otherwise $\mathbf{Y}(t,j)=0$. Likewise, we use $\mathbf{R} \in \mathbb{R}^{M \times N}$ to denote the user-item interactions. The $t$-th group $g_t \in \mathcal{G}$ consists of a set of user members $\mathcal{G}_t = \{ u_1, u_2, ..., u_s, ..., u_{|\mathcal{G}_t|} \}$ where $u_s \in \mathcal{U}$ and $|\mathcal{G}_t|$ is the size of $\mathcal{G}_t$. We denote the interaction set of $g_t$ as $\mathcal{Y}_t = \{ i_1, i_2, ..., i_j, ..., i_{|\mathcal{Y}_t|} \}$ where $i_j \in \mathcal{I}$ and $|\mathcal{Y}_t|$ is the size of $\mathcal{Y}_t$. Then, given a target group $g_t$, the group recommendation task is defined as recommending items that $g_t$ may be interested in.

\subsection{Hypergraph}
Different from simple graphs, a hypergraph is a more general topological structure where the edge (namely hyperedge in the hypergraph) could connect two or more nodes \cite{IHGNN, Wang2020NextitemRW, Yu2021SelfSupervisedMH}. Formally, we define the hypergraph as $G=(\mathcal{V}, \mathcal{E}, \mathbf{H})$ where $\mathcal{V}$ is the vertex set, $\mathcal{E}$ is the hyperedge set, and $\mathbf{H} \in \mathbb{R}^{|\mathcal{V}| \times |\mathcal{E}|}$ depicts the connectivity of the hypergraph as $\mathbf{H}(v,e)=1$ if the hyperedge $e$ connects the vertex $v$, otherwise $\mathbf{H}(v,e)=0$. On this basis, we further give some notations in $G$. Let $\mathcal{E}_v$ denote a set of related hyperedges that connect to the node $v (i.e., \mathcal{E}_v = \{ e \in \mathcal{E} | \mathbf{H}(v,e)=1 \})$ and  $\mathcal{V}_e$ denote a set of nodes to which the hyperedge $e$ connects (i.e., $\mathcal{V}_e = \{ v \in \mathcal{V} | \mathbf{H}(v,e)=1 \}$).

\section{Methodology}
In this section, we first elaborate on the multi-view modeling of group consensus. Then, we introduce their specific encoding process enhanced by graph neural networks. We move on to introduce the adaptive fusion mechanism for integrating multi-view information. Finally, we demonstrate our optimization approach. The overall architecture of ConsRec is shown in Figure \ref{fig:overview}. Notations used in this section are also summarized in Table \ref{tab:notation} in the Appendix.

\subsection{Multi-view Modeling of Consensus}
To effectively exploit the group behavior data for capturing group consensus, we design three novel views which provide complementary information to enable multi-view learning. The illustrative example is shown in the left part of Figure \ref{fig:overview}.

\subsubsection{Member-level} Since groups consist of diverse members, we first analyze each group from a member constitution view, which provides a fine-grained explanation for group decisions. However, conventional graph structure only supports \textit{pairwise} relationship that reflects one-to-one relation. Therefore, we propose to model the \textit{tuplewise} relationship between groups and members with a hypergraph. Formally, we construct the member view with a hypergraph $G^m = (\mathcal{V}^m, \mathcal{E}^m, \mathbf{H}^m)$ where $\mathcal{V}^m = \mathcal{U} \cup \mathcal{I}$ denotes the node set, $\mathcal{E}^m = \mathcal{G}$ denotes the hyperedge set, and the adjacency matrix $\mathbf{H}^m \in \mathbb{R}^{ |\mathcal{V}^m| \times |\mathcal{E}^m|}$ denotes the affiliations among nodes and edges. For the $t$-th group $g_t$, we represent it as the $t$-th hyperedge $e_t \in \mathcal{E}^m$ and connect it to corresponding member nodes as well as item nodes, i.e., $\mathbf{H}^m(s,t)=1$ if node $v_s \in \mathcal{G}_t \cup \mathcal{Y}_t$ where $\mathcal{G}_t$ and $\mathcal{Y}_t$ refer to $g_t$'s member set and interaction set, respectively. For example, in Figure \ref{fig:overview}, as group $g_1$ consists of $u_1$ and $u_2$, and have interacted with $i_1$ and $i_2$, we would represent group $g_1$ with the hyperedge $e_1$ that connects to its member and item nodes. 

\subsubsection{Item-level} Sometimes, the consensus of groups can be inferred from their interacted items. For example, in Figure \ref{fig:example}, we can conclude the group's agreement towards romantic films based on its interaction history. Therefore, to capture the groups' item-level tastes, we construct the item-level view where groups and their interacted items form a bipartite graph. Specifically, we design the item view with the graph $G^i = (\mathcal{V}^i, \mathcal{E}^i, \mathbf{A}^i)$ where $\mathcal{V}^i = \mathcal{G} \cup \mathcal{I}$ denotes the node set, $\mathcal{E}^i$ denotes the edge set as $\mathcal{E}^i = \{ (g_t, i_j) | g_t \in \mathcal{G}, i_j \in \mathcal{I}, \mathbf{Y}(t,j)=1 \} $, and $\mathbf{A}^i \in \mathbb{R}^{(K+N) \times (K+N)}$ is the adjacency matrix as $\mathbf{A}^i = \begin{bmatrix}\mathbf{0} \ \ \ \ \mathbf{Y} \\ \mathbf{Y}^{\text{T}} \ \ \mathbf{0} \end{bmatrix}$. As shown in Figure \ref{fig:overview}, we connect node $g_2$ and $i_3$ since group $g_2$ has watched the movie $i_3$.

\subsubsection{Group-level} In real-world scenarios, groups may have their inherent preferences. For example, a group formed by a family tend to watch family-style comedies. To capture and propagate such signals among similar groups, we devise the group-level view where otherwise isolated groups are connected. Specifically, each group is modeled as a single node and different groups are connected if they share at least a common member or item. Formally, the group-level graph is represented as $G^g=(\mathcal{V}^g, \mathcal{E}^g, \mathbf{A}^g)$ where $\mathcal{V}^g = \mathcal{G}$ and $\mathcal{E}^g = \{(g_p, g_q)| g_p, g_q \in \mathcal{G}, |\mathcal{G}_p \cap \mathcal{G}_q| \geq 1 \text{or}  \; |\mathcal{Y}_p \cap \mathcal{Y}_q| \geq 1 \}$. We utilize the adjacency matrix $\mathbf{A}^g$ to discriminate relevance between groups as assigning each edge $(g_p, g_q)$ with a weight $\mathbf{A}^g(p,q)$ = $\frac{|\mathcal{G}_p \cap \mathcal{G}_q| + |\mathcal{Y}_p \cap \mathcal{Y}_q|}{|\mathcal{G}_p \cup \mathcal{G}_q|+|\mathcal{Y}_p \cup \mathcal{Y}_q|}$. For example, as depicted in Figure \ref{fig:overview}, since $g_1$ and $g_2$ share a common member, we connect these two groups and compute the weight as $\frac{1+0}{3+3}=\frac{1}{6}$.

\subsubsection{Multi-view Modeling} In this paper, we aim to embed users, items, and groups to a lower-dimension vector representation. Based on the above multi-view modeling of group consensus, we can represent the learned $d$-dimensional embedding vectors of users, items, and groups as matrices $\mathbf{U} \in \mathbb{R}^{M \times d}$, $\mathbf{I} \in \mathbb{R}^{N \times d}$, and $\mathbf{G} \in \mathbb{R}^{K \times d}$, respectively. Such matrices can be learned from the following graph representation learning techniques.

\begin{figure*}[!t]
    \centering
    \includegraphics[width=17cm]{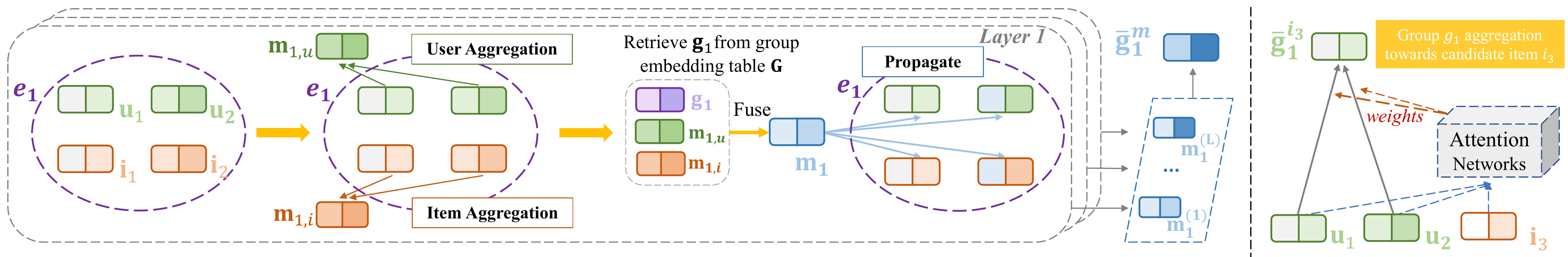}
    \caption{Comparison between our hypergraph learning-based aggregation (left) and the commonly adopted attentive aggregation (right). Ours wins in efficiency, fairness, and expressiveness with details explained in Section \ref{sec:agg_comp}.}
    \label{fig:aggregation_final}
\end{figure*}

\subsection{Member-level Hypergraph Networks}

In this subsection, we introduce the member-level view where we aim to obtain a meaningful member-level aggregation for estimating the consensus. Specifically, we devise a novel preference-aware hypergraph neural network that realizes an efficient, fair, and expressive aggregation schema.

\subsubsection{Preference-aware hypergraph neural network (P-HGNN)} 
We feed user embeddings $\mathbf{U} \in \mathbb{R}^{M \times d}$ and item embeddings $\mathbf{I} \in \mathbb{R}^{N \times d}$ to the hypergraph neural network. Meanwhile, since we represent each group as a hyperedge $e$, its embedding can be retrieved from the group embedding table $\mathbf{G}$ as $\mathbf{g}_e = \mathbf{G}(e,:)$. Technically, our goal is to obtain refined item embeddings $\overline{\mathbf{I}}$ and groups' member-level representations $\overline{\mathbf{G}}^m$ via hypergraph propagation. We move on to introduce the specific design of this hypergraph neural network with the illustrative process shown in Figure \ref{fig:aggregation_final}.

\noindent{\textbf{Efficiency.}} In hypergraph neural networks \cite{HyperGCN, HNN}, hyperedges serve as mediums for processing and transferring information. Recall that we represent each group as a hyperedge in the member-level hypergraph, it is straightforward to regard the carried messages of hyperedges during convolution as the member-level aggregation. In this way, member-level aggregation can be acquired much more efficiently compared with the exhaustive and candidate-reliant attentive aggregation adopted by \cite{AGREE, SoAGREE, GAME, HCR, HHGR}. 

Formally, a hyperedge $e \in \mathcal{E}^m$ connects both member nodes and item nodes which preserve different semantic information. Therefore, we separate the item aggregation and user aggregation to maintain their distinction. Specifically, for hyperedge $e$, we compute its member aggregated message $\mathbf{m}_{e,u}$ within its member set $\mathcal{G}_e$ as $\mathbf{m}_{e,u} = \textbf{AGG}_{node} ( \{ \mathbf{u}_s | u_s \in \mathcal{G}_e \})$. Here $\textbf{AGG}_{node}(\cdot)$ denotes a specific node-aggregation function and $\mathbf{u}_s \in \mathbb{R}^d$ denotes the embedding of $s$-th user as $\mathbf{u}_s = \mathbf{U}(s,:)$. Likewise, we obtain $e$'s item-side message $\mathbf{m}_{e,i}$ within its interaction set $\mathcal{Y}_e$ as $\mathbf{m}_{e,i} = \textbf{AGG}_{node}(\{ \mathbf{i}_j | i_j \in \mathcal{Y}_e  \})$ where $\mathbf{i}_j$ denotes the embedding of $j$-th item as $\mathbf{i}_j = \mathbf{I}(j,:)$. Then, it is intuitive to utilize these messages to synthesize the hyperedge $e$'s representation $\mathbf{m}_e$.

\noindent{\textbf{Fairness.}} As shown in Figure \ref{fig:aggregation_final}, the commonly adopted attentive aggregation assigns more weights to active members and thus obtain an unfair aggregation \cite{AttentionBias}. Hopefully, this problem can be alleviated as we introduce the item-side messages as a supplementary. Furthermore, we utilize the hyperedge $e$'s independent representation $\mathbf{g}_e$ to interact with the item messages $\mathbf{m}_{e,i}$ for generating extra collaborative signals. Then, we fuse these different information via linear transformation as follows: 
\begin{equation}
    \mathbf{m}_e = \textbf{CONCAT}(\mathbf{m}_{e,u} \;, \; \mathbf{m}_{e,i} \;, \; \mathbf{m}_{e,i} \odot \mathbf{g}_e) \mathbf{W}^f, 
    \label{eq: fusion}
\end{equation}

\noindent where $\mathbf{m}_e$ denotes the synthesized messages carried by $e$, $\odot$ stands for element-wise product, and $\mathbf{W}^f \in \mathbb{R}^{3d \times d}$ denotes the trainable weight matrix for messages fusion. After applying this operation, the messages contained by hyperedges are more meaningful and unbiased as revealing the common information between members and groups. Finally, we can refine the representations of node $v$ by collecting the messages from its related hyperedges. Formally, for a target item node $i_j$, its representation is updated as:
\begin{equation}
    \small
    \overline{\mathbf{i}}_j = \textbf{AGG}_{he}(\{ \mathbf{m}_e | e \in \mathcal{E}_j \}),
    \label{eq:item_update}
\end{equation}
\noindent where $\overline{\mathbf{i}}_j$ denotes the refined embedding of node $i_j$, $\textbf{AGG}_{he}(\cdot)$ denotes a specific hyperedge-aggregation function, and $\mathcal{E}_j$ is the set of hyperedges that connect to node $i_j$. So far, we have accomplished the goal of obtaining refined item embeddings $\overline{\mathbf{I}}$ and groups' member-level representations $\overline{\mathbf{G}}^m$ by stacking all $\overline{\mathbf{i}}_j \;(j=1,...,N)$ and $\mathbf{m}_e \;(e=1,...,K)$, respectively.

\noindent{\textbf{Expressiveness.}} To improve expressiveness, we further stack aforementioned propagation module for multiple layers so that both nodes' representations and hyperedges' messages can benefit from high-order neighbors. Finally, we average the embedding obtained at each layer to generate the final representation for node $i_j$ as: 

\begin{equation*}
    \overline{\mathbf{i}}_j = \frac{1}{L+1} \sum_{l=0}^L \mathbf{i}_j^{(l)},
\end{equation*}
\noindent where $L$ is the total number of convolutional layers, $\mathbf{i}_j^{(l)}$ is the representation of node $i_j$ in the $l$-th layer (we have $\mathbf{i}_j^{(0)} = \mathbf{i}_j$ and the calculation of subsequent layers can follow Equations \ref{eq: fusion} and \ref{eq:item_update}). Similarly, we average the aggregated messages during each layer to generate the final representation for hyperedge (group) $e$ as: 

\begin{equation*}
    \overline{\mathbf{g}}_e^m = \frac{1}{L+1}  \sum_{l=0}^L \mathbf{m}_e^{(l)},
\end{equation*}

\noindent where $\overline{\mathbf{g}}^m_e$ denotes the member-level preferences of group $e$, and $\mathbf{m}_e^{(l)}$ denotes the messages contained by $e$ in the $l$-th layer which can be calculated by Equation \ref{eq: fusion}. By stacking all the $\overline{\mathbf{g}}_e^m \;(e=1,...,K)$, we can get the groups' member-level representations $\overline{\mathbf{G}}^m$. Besides, we implement $\textbf{AGG}_{node}(\cdot)$ and $\textbf{AGG}_{he}(\cdot)$ with the average pooling due to its simplicity and effectiveness.

\subsubsection{Discussions\label{sec:agg_comp}} We compare our aggregation with existing attentive aggregation adopted by \cite{AGREE, SoAGREE, GAME, HyperGroup, HCR, HHGR}. The illustrative example is shown in Figure \ref{fig:aggregation_final}. Given a candidate item, the attentive method first computes the attention weights by feeding the concatenation of member and item representations to an attention network, then sums over members based on the weights to obtain the group's representation. Considering the huge amount of candidate items in real-world scenes, this candidate-reliant method is infeasible. Besides, this mode is prone to assigning active members with more weights, leading to a biased aggregation. On the contrary, our aggregation schema wins in efficiency as obtaining aggregation via hypergraph convolutional operations, fairness as introducing side information for supplementary, and expressiveness as incorporating high-order collaborative information.

\subsection{Item-level Graph Networks}
Recall that we design the group-item bipartite graph $G^i$ to model the consensus from item-level interests. In this subsection, we aim to employ the graph neural networks \cite{GCN, LightGCN} to capture collaborative signals between groups and items, finally obtaining the groups' discriminative representations at the item-level. 

We feed the concatenation of group embeddings $\mathbf{G} \in \mathbb{R}^{K \times d}$ and item embeddings $\mathbf{I} \in \mathbb{R}^{N \times d}$ to the graph convolutional network, denoted as $\mathbf{E}=\begin{bmatrix} \mathbf{G} \\ \mathbf{I} \end{bmatrix}$. Referring to the spectral graph convolution \cite{GCN, LightGCN, SimplifyingGNN}, we define our graph convolution in the $l$-th layer as:

\begin{equation}
    \mathbf{E}^{(l+1)} = \mathbf{D}^{-\frac{1}{2}} \mathbf{A}^i  \mathbf{D}^{-\frac{1}{2}} \mathbf{E}^{(l)},
    \label{eq:gnn}
\end{equation}

\noindent where $\mathbf{E}^{(l)}$ denotes the node representations in the $l$-th layer and $\mathbf{D}$ is the diagonal node degree matrix of adjacency matrix $\mathbf{A}^i$. We pass $\mathbf{E}^{(0)} = \mathbf{E}$ through $L$ convolutional layers, and then average the embeddings obtained at each layer to get the final embeddings as $\overline{\mathbf{E}} = \frac{1}{L+1} \sum_{l=0} ^{L} \mathbf{E}^{(l)} = \begin{bmatrix} \overline{\mathbf{G}}^i \\  \overline{\mathbf{I}}^i \end{bmatrix}$. Therefore, group $e$'s item-level consensus can be obtained as $\overline{\mathbf{g}}_e^i = \overline{\mathbf{G}}^i(e,:)$. As shown in Figure \ref{fig:overview}, during propagation, interacted items' information has been explicitly injected to groups' item-level representations, and thus the obtained $\overline{\mathbf{G}}^i$ is expressive enough to reflect item-level tastes.

\subsection{Group-level Graph Networks}
Since the constructed group-level graph $G^g$ depicts the connectivity between groups, we still apply graph neural networks to encode the high-order relations among different groups. Note that we distinguish the relevance between groups by computing different weights, allowing each group to enrich its expressivity by absorbing information from the most relevant neighbors.

Specifically, we feed the group embeddings $\mathbf{G} \in \mathbb{R}^{K \times d}$ to the graph convolutional network, denoted as $\mathbf{G}^{(0)} = \mathbf{G}$. Then the propagation mechanism at each layer is similar to Equation \ref{eq:gnn}. Therefore, after passing $\mathbf{G}^{(0)}$ through $L$ convolutional layers, we can obtain the final group-level embeddings as $\overline{\mathbf{G}}^g = \frac{1}{L+1} \sum_{l=0}^{L} \mathbf{G}^{(l)}$. We regard $\overline{\mathbf{G}}^g$ as group-level inherent preferences as it explores and preserves the group-level proximity.

\subsection{Adaptive Fusion \& Optimization}
As mentioned before, we have modeled the groups' consensus from three different views, including member-level aggregation $\overline{\mathbf{G}}^m$, item-level interests $\overline{\mathbf{G}}^i$, and group-level inherent preferences $\overline{\mathbf{G}}^g$. Then the core problem takes down to fuse them to obtain the final representations of group consensus. To adaptively control the combination of three view-specific group representations, we propose to employ three different gates as:

\begin{equation}
  \label{eq:fuse_func}
     \overline{\mathbf{G}} = \bm{\alpha}  \overline{\mathbf{G}}^m + \bm{\beta}  \overline{\mathbf{G}}^i+ \bm{\gamma}  \overline{\mathbf{G}}^g, 
\end{equation}
\noindent where $\bm{\alpha} = \sigma(\overline{\mathbf{G}}^m  \mathbf{W}^m)$, $\bm{\beta} = \sigma(\overline{\mathbf{G}}^i  \mathbf{W}^i)$, and $\bm{\gamma} = \sigma(\overline{\mathbf{G}}^g  \mathbf{W}^g)$. Here, $\mathbf{W}^m, \mathbf{W}^i$, and $\mathbf{W}^g \in \mathbb{R}^{d }$ are three different trainable weights and $\sigma$ is the activation function. The $\bm{\alpha}, \bm{\beta}$, and $\bm{\gamma}$ denote the learned weights to balance the contributions of member-level, item-level, and group-level information, respectively. On this basis, we can obtain the final group representations $\overline{\mathbf{G}}$. During optimization, based on the abundant supervision signals, our model can automatically discriminate the importance of different views.

Then, we introduce our optimization strategy that jointly learns user-item and group-item interactions. Detailed training procedure is shown in Algorithm \ref{alg: train}. For the group-item pair $(g_t, i_j)$, we feed their corresponding representations to a Multi-layer Perceptron (MLP) \cite{MLP} to compute the final prediction score $\hat{y}_{t,j}$ as $\hat{y}_{tj} = \textbf{MLP}(\overline{\mathbf{g}}_t \odot \overline{\mathbf{i}}_j),$ where $\overline{\mathbf{g}}_t = \overline{\mathbf{G}}(t,:)$ denotes the final group representation of the target group $g_t$, $\overline{\mathbf{i}}_j = \overline{\mathbf{I}}(j,:)$ denotes the refined embedding of the candidate item $i_j$, and $\textbf{MLP}(\cdot)$ is designed with input dimension set as $d$ and output dimension as 1. With the group-item interaction data, we utilize the \textit{Bayesian Personalized Ranking} (BPR) loss \cite{BPRloss} for optimization as follows:

\begin{equation}
    \mathcal{L}_{group} = - \sum_{g_t \in \mathcal{G}} \frac{1}{|\mathcal{D}_{g_t}|} \sum_{(j, j') \in \mathcal{D}_{g_t} } \text{ln} \; \sigma (\hat{y}_{tj} - \hat{y}_{tj'}),
    \label{eq:group_loss}
\end{equation}

\noindent where $\mathcal{D}_{g_t}$ denotes the group-item training set sampled for group $g_t$, in which each instance is a pair $(j, j')$ meaning that group $g_t$ has interacted with item $i_j$, but has not interacted with item $i_{j'}$.

To further utilize supervision signals, we propose to incorporate the user-item interaction data to optimize the group-item and user-item tasks simultaneously. Similarly, for the user-item pair $(u_s, i_j)$, we compute the prediction score $\hat{r}_{sj} = \textbf{MLP}(\mathbf{u}_s \odot \mathbf{i}_j)$, where the $\textbf{MLP}(\cdot)$ is shared with the group-item MLP network, and $\mathbf{u}_s$ and $\mathbf{i}_j$ denote the corresponding user and item embedding. We utilize the same pairwise loss function for optimization as follows:

\begin{equation}
    \mathcal{L}_{user} = - \sum_{u_s \in \mathcal{U}} \frac{1}{|\mathcal{D}_{u_s}|} \sum_{(j, j') \in \mathcal{D}_{u_s}} \text{ln} \; \sigma (\hat{r}_{sj} - \hat{r}_{sj'}), 
    \label{eq:user_loss}
\end{equation}

\noindent where $\mathcal{D}_{u_s}$ denotes the user-item training set sampled for user $u_s$ and $(j, j')$ represents user $u_s$ prefers observed item $i_j$ over unobserved item $i_{j'}$. We jointly train $\mathcal{L}_{group}$ and $\mathcal{L}_{user}$ on all the group-item and user-item interactions as $\mathcal{L} = \mathcal{L}_{group} + \mathcal{L}_{user}$. 


\section{Experiment}
In this section, we present our experimental setup and empirical results. Our experiments are designed to answer the following research questions (RQs):
\begin{itemize}[leftmargin=*, topsep=0pt]
    \item \textbf{RQ1:} How does our proposed ConsRec perform compared with various group recommendation methods?
    \item \textbf{RQ2:} Whether the proposed different views capture the consensus information and provide better recommendation?
    \item \textbf{RQ3:} How efficient is our method compared with other group recommendation techniques?
    \item \textbf{RQ4:} How do our preference-aware hypergraph neural network (P-HGNN) work?
    \item \textbf{RQ5:} How do different settings of hyperparameters affect the model performance?
\end{itemize}
Due to space limitation, we move the RQ4 and 5 to the Appendix.

 \subsection{Experimental Settings}
 \subsubsection{Datasets.} We conduct experiments on two real-world public datasets, Mafengwo and CAMRa2011 \cite{AGREE}. Specifically, Mafengwo is a tourism website where users can record their traveled venues, create or join a group travel. CAMRa2011 is a real-world dataset containing the movie rating records of individual users and households. Table \ref{table:datasets} reports the detailed statistics of experimental datasets.

 \begin{table}[!t]
 \small
  \caption{Statistics of datasets.}
  \label{table:datasets}
  \begin{tabular}{c|ccccc}
    \toprule
     \multirow{2}{*}{Dataset} & \multirow{2}{*}{$\#$Users} & \multirow{2}{*}{$\#$Items} & \multirow{2}{*}{$\#$Groups} &  $\#$U-I & $\#$G-I \\
     & & & & interactions & interactions \\
    \midrule
    Mafengwo & 5,275 & 1,513 & 995 & 39,761 & 3,595 \\
    CAMRa2011 & 602 & 7,710 & 290 & 116,344 & 145,068 \\
  
  \bottomrule
\end{tabular}
\end{table}

\subsubsection{Baselines.} To evaluate the effectiveness of ConsRec, we compare it with the following representative approaches. Due to space limitation, we only show their belonging categories and move the details to Appendix \ref{sec:baseline}. Note that these categories have overlaps. For example, the baseline $\mathbf{S}^2$-HHGR \cite{HHGR} integrates both hypergraph learning and self-supervised learning. 

\begin{itemize}[leftmargin=*, topsep=0pt]
    \item \textbf{Non-personalized method}: Popularity (Pop) \cite{Popularity}
    \item \textbf{Classical neural network based method}: NCF \cite{NCF} 
    \item \textbf{Attentive aggregation method}: AGREE \cite{AGREE}
    \item \textbf{Hypergraph learning-enhanced}: HyperGroup \cite{HyperGroup}, HCR \cite{HCR}
    \item \textbf{Self-supervised learning-enhanced}: GroupIM \cite{GroupIM}, $\mathbf{S}^2$-HHGR \cite{HHGR}, and CubeRec \cite{HyperCube}
\end{itemize}

\begin{table*}[t]
 \small
  \caption{Performance comparison of all methods on group recommendation task in terms of HR@K and NDCG@K.}
  \label{table:group_rec}
  \begin{tabular}{c|c|ccccccccc}
    \toprule
    Dataset & Metric & Pop & NCF & AGREE & HyperGroup & HCR & GroupIM & $\mathbf{S}^2$-HHGR  & CubeRec & \textbf{ConsRec} \\
      \midrule
    \multirow{4}{*}{Mafengwo } & HR@5 & 0.3115 & 0.4701 & 0.4729 & 0.5739 & 0.7759 & 0.7377 & 0.7568 & \underline{0.8613} & \textbf{0.8844} \\ 
    & HR@10 & 0.4251 & 0.6269 & 0.6321 & 0.6482 & 0.8503 & 0.8161 & 0.7779 & \underline{0.9025} & \textbf{0.9156} \\ 
    & NDCG@5 & 0.2169 & 0.3657 & 0.3694 & 0.4777 & 0.6611 & 0.6078 & 0.7322 & \underline{0.7574} & \textbf{0.7692} \\ 
    & NDCG@10 & 0.2537 & 0.4141 & 0.4203 & 0.5018 & 0.6852 & 0.6330 & 0.7391 & \underline{0.7708} & \textbf{0.7794} \\ 
    \midrule
    
    \multirow{4}{*}{CAMRa2011} & HR@5 & 0.4324 & 0.5803 & 0.5879 & 0.5890 & 0.5883  & \textbf{0.6552} & 0.6062 & 0.6400 & \underline{0.6407} \\ 
    & HR@10 & 0.5793 & 0.7693 & 0.7789 & 0.7986 & 0.7821 & \textbf{0.8407} & 0.7903 & 0.8207 & \underline{0.8248} \\
    & NDCG@5 & 0.2825 & 0.3896 & 0.3933 & 0.3856 & 0.4044 & 0.4310  & 0.3853 & \underline{0.4346} & \textbf{0.4358} \\ 
    & NDCG@10 & 0.3302 & 0.4448 & 0.4530 & 0.4538 & 0.4670 & 0.4914 & 0.4453 & \underline{0.4935} & \textbf{0.4945} \\
  \bottomrule
\end{tabular}
\end{table*}

\begin{table*}[t]
 \small
  \caption{Performance comparison of all methods on user recommendation task in terms of HR@K and NDCG@K.}
  \label{table:user_rec}
  \begin{tabular}{c|c|ccccccccc}
    \toprule
    Dataset & Metric & Pop & NCF & AGREE & HyperGroup & HCR & GroupIM & $\mathbf{S}^2$-HHGR & CubeRec & \textbf{ConsRec} \\
      \midrule
    \multirow{4}{*}{Mafengwo } & HR@5 & 0.4047 & 0.6363 & 0.6357 & 0.7235 & \underline{0.7571} & 0.1608  & 0.6380 & 0.1847 & \textbf{0.7725} \\
    
    & HR@10 & 0.4971 & 0.7417 & 0.7403 & 0.7759 & \underline{0.8290} & 0.2497 & 0.7520 & 0.3734 & \textbf{0.8404} \\ 
    & NDCG@5 & 0.2876 & 0.5432 & 0.5481 & 0.6722  & \underline{0.6703} & 0.1134 & 0.4637 & 0.1099 & \textbf{0.6884} \\ 
    & NDCG@10 & 0.3172 & 0.5733 & 0.5738 & 0.6894 & \underline{0.6937} & 0.1420 & 0.5006 & 0.1708 & \textbf{0.7107} \\

    \midrule
    \multirow{4}{*}{CAMRa2011} & HR@5 & 0.4624 & 0.6119 & 0.6196 & 0.5728 & \underline{0.6262} & 0.6113 & 0.6153 & 0.5754  & \textbf{0.6774}\\
    & HR@10 & 0.6026 & 0.7894 & 0.7897 & 0.7601 & 0.7924 & 0.7771  & \underline{0.8173} & 0.7827 & \textbf{0.8412} \\
    & NDCG@5 & 0.3104 & 0.4018 & 0.4098 & \underline{0.4410} & 0.4195 & 0.4064& 0.3978 & 0.3751 & \textbf{0.4568} \\ 
    & NDCG@10 & 0.3560 & 0.4535 & 0.4627 & \underline{0.5016} & 0.4734 &  0.4606& 0.4641 & 0.4428 & \textbf{0.5104} \\ 
  \bottomrule
\end{tabular}
\end{table*}

\subsubsection{Evaluation Metrics.} Following previous settings \cite{AGREE, SoAGREE, HCR}, we adopt two widely used evaluation metrics in terms of top-K recommendation, \textit{i.e.}, Hit Ratio (HR) and Normalized Discounted Cumulative Gain (NDCG). Higher HR and NDCG indicate better performance. To alleviate the heavy computation with all items serving as the candidates, we randomly sample 100 negative items for each ground-truth item and rank them according to the calculated interaction probabilities \cite{AGREE, HCR}. We evaluate the performance of all methods over the same metrics and test data. Due to space limitation, data pre-processing and implementation details are moved to Appendix \ref{sec:implement}.

\subsection{Overall Performance (RQ1)}
 We compare the performance of ConsRec with all baselines. Tables \ref{table:group_rec} and \ref{table:user_rec} show the experimental performance on group recommendation task and user recommendation task, respectively. According to the results, we note the following key observations:

\begin{itemize}[leftmargin=*, topsep=5pt]
    \item For the group recommendation task shown in Table \ref{table:group_rec}, ConsRec outperforms all baselines under most evaluation metrics on two benchmark datasets. We own our superiority to the proposed three novel views of consensus modeling, capturing the preferences of groups in the most comprehensive manner.

    \item Among all baselines, hypergraph learning-enhanced models (HyperGroup \cite{HyperGroup} and HCR \cite{HCR}) achieve better performance than classical aggregation method (AGREE \cite{AGREE}), indicating the necessity of capturing high-order collaborative information. Besides, $\mathbf{S}^2$-HHGR \cite{HHGR} and CubeRec \cite{HyperCube} achieve further improvement due to the introduction of self-supervised learning objectives. Nonetheless, our model realizes the best performance.

    \item For the user recommendation task shown in Table \ref{table:user_rec}, our model still has advantages. We attribute this to our optimization strategy that boosts group and user recommendation simultaneously. It is worth noting that GroupIM \cite{GroupIM} and CubeRec \cite{HyperCube}, though achieving ideal performance on group recommendation, perform poorly on user recommendation. This is because they separate user-item and group-item training, sacrificing the performance of user recommendation to overfit the group recommendation.
\end{itemize}

\begin{table}[t]
    \centering
    \caption{Ablation study on different views with group recommendation results reported. ``w/o. M'', ``w/o. I'', and ``w/o. G'' refer to the variant that eliminates the member-level, item-item, and group-level view, respectively.}
    \small
    \begin{tabular}{c|c|ccc|c}
    \toprule
        Dataset & Metric & w/o. M & w/o. I & w/o. G & \textbf{Full}  \\
        \midrule
      \multirow{4}{*}{Mafengwo} & HR@5 & 0.8201 & 0.8704 & 0.8593 & \textbf{0.8844}     \\ 
      & HR@10 & 0.8724 & 0.9075 & 0.9005 & \textbf{0.9156} \\ 
      & NDCG@5 & 0.7021 & 0.7597 & 0.7376 & \textbf{0.7692} \\
      & NDCG@10 & 0.7192 & 0.7718 & 0. 7510 & \textbf{0.7794} \\ 
      
      \bottomrule
    \end{tabular}
    \label{tab:different_view}
    \vspace{-2.0em}
\end{table}

\subsection{Effectiveness of Consensus Learning (RQ2)}
\subsubsection{Multi-view Modeling} To capture the group consensus, we innovatively devise three views, including member-level aggregation, item-level interests, and group-level inherent preferences. To verify whether each view has captured distinct aspect of group information, we further conduct ablation study. Technically, at each time, we remove one single view and only fuse the remaining two views following the same adaptive fusion mechanism as Equation \ref{eq:fuse_func}. These three variants are denoted as ``w/o. M'', ``w/o. I'', and ``w/o. G'', respectively. We report the experimental results in Table \ref{tab:different_view}. Due to space limitations, we only list the results on Mafengwo as CAMRa2011 shares a similar performance pattern. From this table, we can observe that removing any view degrades the performance, showing that each view plays a distinct role in capturing the consensus.

\subsubsection{Case Study}
\begin{figure*}[!t]
    \centering
    \includegraphics[width=18cm]{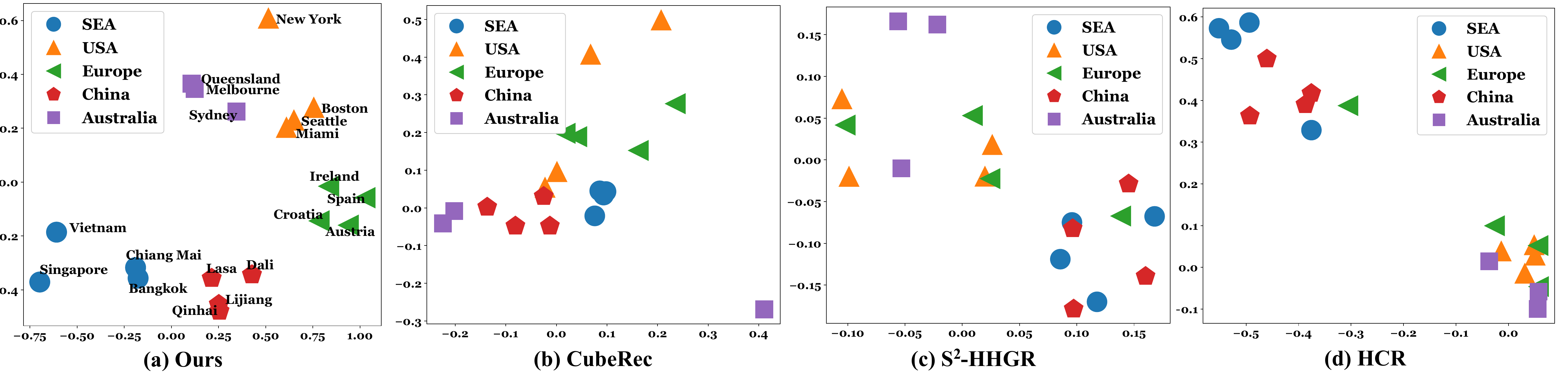}
    \caption{Visualization of learned item embeddings. We plot two dimensions of item representations on Mafengwo-S. ConsRec learns the latent properties of items as geographically similar items are close to each other in the embedding space.}
    \label{fig:case_item}
\end{figure*}

We further conduct a case study to explore how ConsRec captures consensus information on the specific example. However, our two experimental datasets do not contain any semantic information (\textit{e.g.}, item's name or content) to facilitate interpretability. Therefore, we crawl an extra Mafengwo dataset from its official website. This dataset is named ``Mafengwo-S'' where ``S'' refers to semantics. We move the detailed crawling process to the Appendix \ref{sec:Crawl}. Finally, we obtain a new dataset that contains 11,027 users, 1,215 groups, and 1,236 items. Besides, each item has distinct semantic information, \textit{i.e.}, the location. For comparison, we only consider HCR \cite{HCR}, GroupIM \cite{GroupIM}, $\mathbf{S}^2$-HHGR \cite{HHGR}, and CubeRec \cite{HyperCube} since other baselines have been shown inferior performance.

\begin{figure}[!t]
    \centering
    \includegraphics[width=7cm]{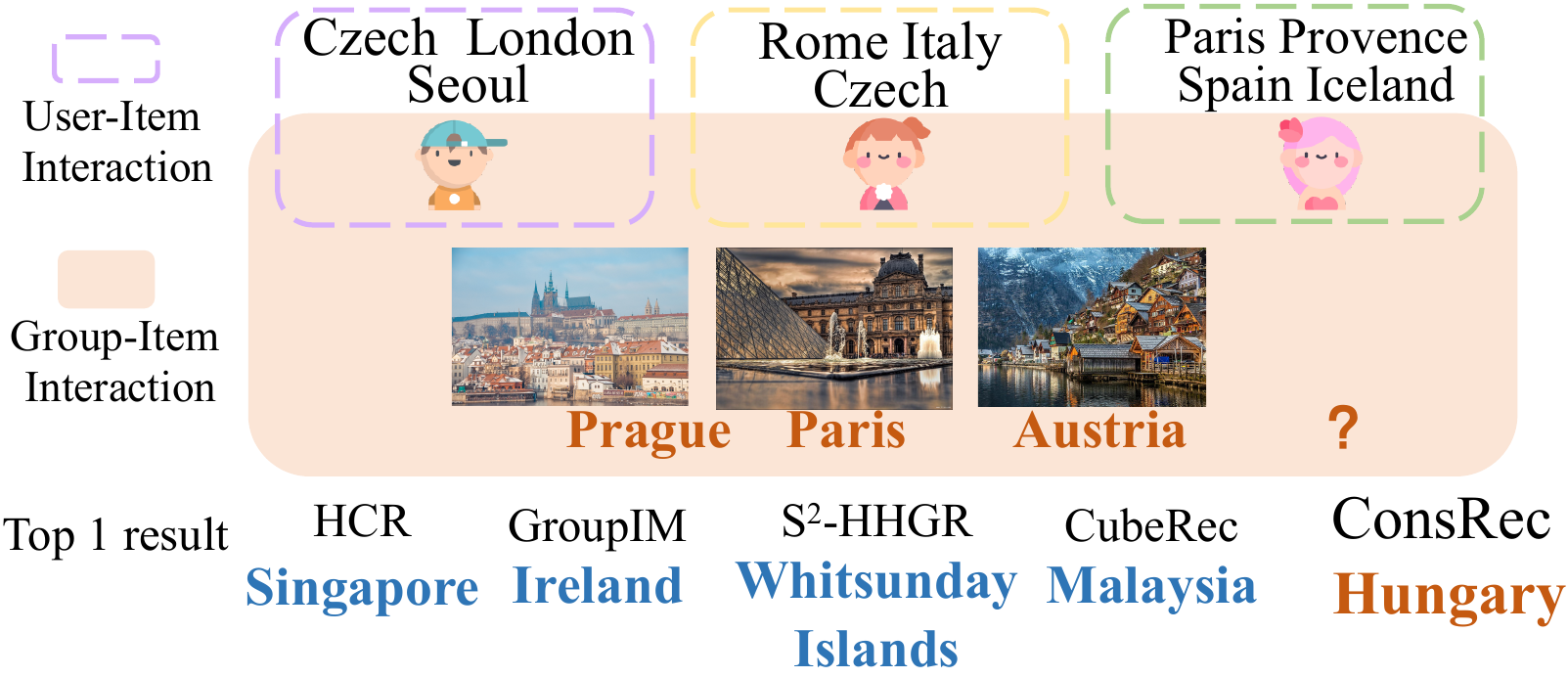}
    \caption{Case study on Mafengwo-S. Both the group and its members have visited European cities. ConsRec captures this consensus and suggests Hungary that hits the ground truth. On the contrary, HCR, GroupIM, $\mathbf{S}^2$-HHGR, and CubeRec are biased by one member’s interests towards Iceland and recommend unsatisfying islands or coastal cities.}
    \label{fig:case}
\end{figure}

 As shown in Figure \ref{fig:case}, this is a group consisting of three members, and both members and the group have visited European cities. ConsRec captures the group's consensus towards European cities and suggests Hungary that hits the ground truth. HCR, GroupIM, $\mathbf{S}^2$-HHGR, and CubeRec are biased by one member's preference toward Iceland and then recommend unsatisfying islands. 

Besides, we also investigate the items' (travel locations) representations learned by different methods. Since GroupIM \cite{GroupIM} do not maintain item embedding tables, we only compare ConsRec with HCR \cite{HCR}, $\mathbf{S}^2$-HHGR \cite{HHGR}, and CubeRec \cite{HyperCube}. Specifically, we select different locations from Southeast Asia (SEA), the USA, Europe, China, and Australia, and visualize their corresponding embeddings in Figure \ref{fig:case_item}. It is worth noting that ConsRec can learn the latent properties of items as geographically similar items are close to each other in the embedding space. On the contrary, the embeddings learned by other baselines are not that discriminative.

We also present the overall performance comparison on the group recommendation task. From Table \ref{tab:mafengwos}, ConsRec consistently ourperforms other strong baselines, showing its superiority.

\begin{table}[!t]
    \centering
    \small
      \caption{Performance comparison on group recommendation task on Mafengwo-S dataset.}
    \begin{tabular}{c|cccc|c}
    \toprule
    Metric & HCR & GroupIM & $\mathbf{S}^2$-HHGR & CubeRec & \textbf{ConsRec} \\
    \midrule
      HR@5   & 0.4845 & 0.5824 & 0.5928 & \underline{0.6237} & \textbf{0.6409} \\
      HR@10   & 0.6099 & \underline{0.6959} & 0.6546 & 0.6873 & \textbf{0.6993} \\

      NDCG@5 & 0.3947 & 0.4591  &  0.5348 & \underline{0.5357} & \textbf{0.5447} \\
      NDCG@10 & 0.4353 & 0.4983  & 0.5545 & \underline{0.5567} & \textbf{0.5642} \\
      
        \bottomrule
    \end{tabular}
    \label{tab:mafengwos}
\end{table}

\subsection{Efficiency Study (RQ3)\label{sec:efficiency}}

\begin{figure}[!t]
    \centering
     \includegraphics[width=8cm]{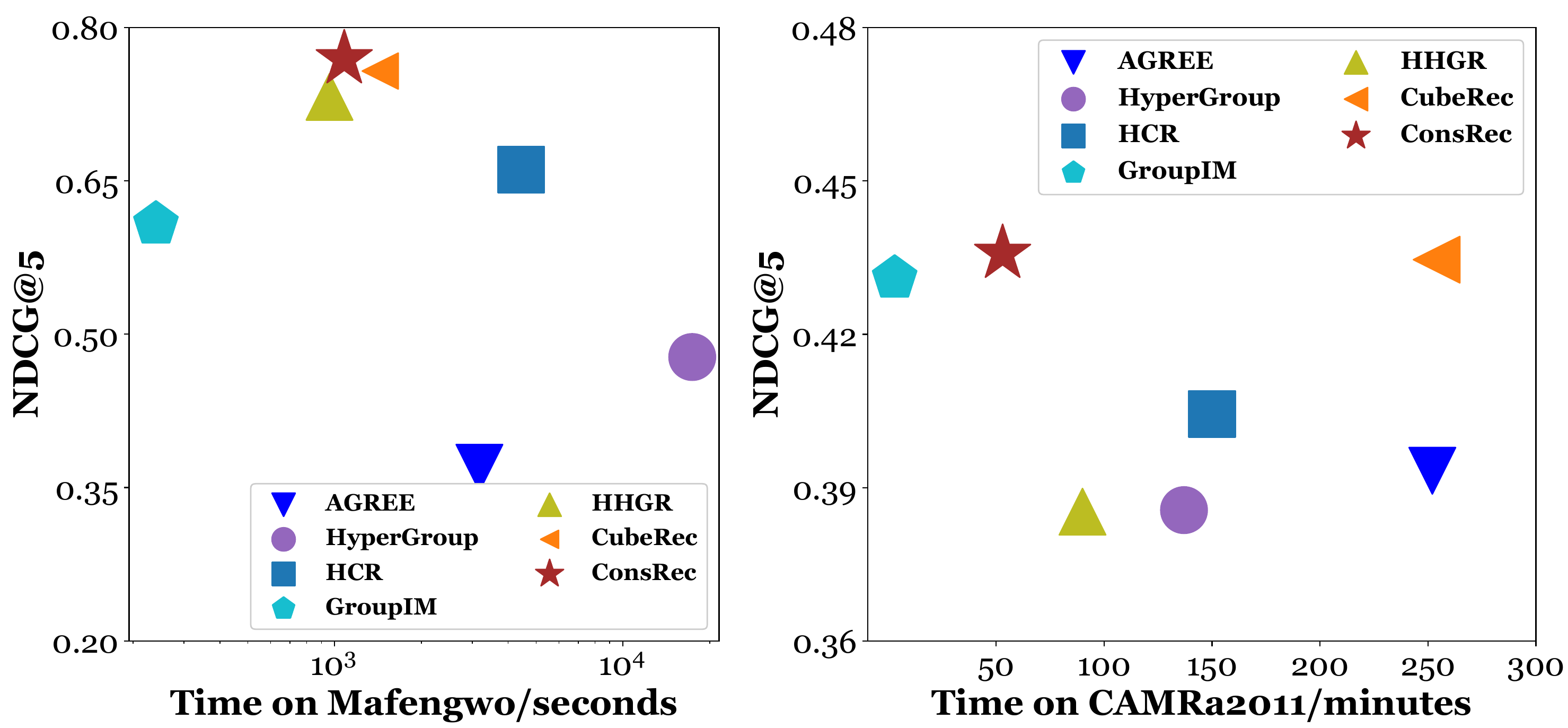}
     \caption{Efficiency Study}
     \label{fig:efficiency}
 \end{figure}

We evaluate the efficiency of ConsRec by directly comparing the total running time (training plus testing) with all baselines. Figure \ref{fig:efficiency} shows the performance (NDCG@5) and running time (seconds or minutes) on two experimental datasets. Notably, our method is quite efficient among various group recommendation baselines and it achieves the best performance simultaneously. On CAMRa2011, our advantages are more obvious. Since CAMRa2011 is much denser than Mafengwo, this phenomenon illustrates our superiority in alleviating computational overhead in larger datasets.

\section{Conclusion}
In this paper, we reveal the consensus behind groups' decision-making and propose a novel recommender ConsRec. To capture the consensus information, ConsRec designs three novel views, including member-level aggregation, item-level tastes, and group-level inherent preferences. Especially, in member-level view, we utilize hypergraph learning to realize an efficient and expressive member-level aggregation. Extensive experiments on two real-world datasets show the effectiveness and efficiency of ConsRec.

\balance
\begin{acks}
This work is also partially supported by the National Natural Science Foundation of China Projects No. U1936213, NSF through grants IIS-1763365, IIS-2106972, III-1763325, III-1909323, III-2106758, and SaTC-1930941.
\end{acks}
\bibliographystyle{ReferenceFormat/ACM-Reference-Format}
\bibliography{reference}

\newpage
\appendix
\section{Notations}
We list the important notations used in this paper in Table \ref{tab:notation}.

\begin{table}[h]
    \centering
    \small
    \caption{Important Notations}
    \begin{tabular}{c|c}
       \toprule
       Symbol  &  Description \\
       \midrule
       
       $ \mathcal{U}, \mathcal{I}, \mathcal{G}$  & Sets of users, items, and groups \\ 
       $M, N, K$ & Numbers of users, items, and groups \\ 
       $u_s, i_j, g_t$ & The $s$-th user, the $j$-th item, and the $t$-th group \\ 
       $\mathbf{Y}$ & Group-item interaction matrix \\ 
       $\mathbf{R}$ & User-item interaction matrix \\ 
       $\mathcal{G}_t$ & Member set of the $t$-th group \\ 
       $\mathcal{Y}_t$ & Interaction set of the $t$-th group \\ 
       $\mathbf{U}, \mathbf{I}, \mathbf{G}$ & Embedding tables of users, items, and groups \\

       \midrule

       $G^m = (\mathcal{V}^m, \mathcal{E}^m, \mathbf{H}^m)$ & Member-level hypergraph \\ 
       $e_t$ & The $t$-th hyperedge in $G^m$ \\ 
       $\mathbf{m}_{e,u}$ & Aggregated user messages of hyperedge $e$ \\ 
       $\mathbf{m}_{e,i}$ & Aggregated item messages of hyperedge $e$ \\ 
       $\mathbf{m}_e^{(l)}$ & Messages of hyperedge $e$ in the $l$-th layer \\ 
       $\overline{\mathbf{G}}^m$ & Group's member-level representations \\ 
       $\overline{\mathbf{g}}_t^m = \overline{\mathbf{G}}^m(t,:)$ & Member-level representation of the $t$-th group \\

       \midrule

       $G^i=(\mathcal{V}^i, \mathcal{E}^i, \mathbf{A}^i)$ & Item-level bipartite graph \\ 

       $\overline{\mathbf{G}}^i$ & Group's item-level representations \\ 
       $\overline{\mathbf{g}}_t^i = \overline{\mathbf{G}}^i(t,:)$ & Item-level representation of the $t$-th group \\

       \midrule

       $G^g = (\mathcal{V}^g, \mathcal{E}^g, \mathbf{A}^g)$ & Group-level graph \\ 
       $\overline{\mathbf{G}}^g$ & Group-level representations \\ 
       $\overline{\mathbf{g}}_t^g = \overline{\mathbf{G}}^g(t,:)$ & Group-level representation of the $t$-th group  \\
       \midrule

       $\overline{\mathbf{I}}$ & Refined item embeddings \\ 
       $\overline{\mathbf{G}}$ & Final group embeddings \\
    
       \bottomrule
        
    \end{tabular}
    \label{tab:notation}
\end{table}

\begin{algorithm}
\small
  \caption{Training Procedure of ConsRec}
  \label{alg: train}
  \KwInput{Sets of users $\mathcal{U}$, items $\mathcal{I}$, and groups $\mathcal{G}$, group-item interaction matrix $\mathbf{Y}$, and user-item interaction matrix $\mathbf{R}$}
 \KwOutput{All model parameters collectively referred to as $\Theta$}

  Randomly initialize $\Theta$; 

  \While{\text{not converge}}
  {
      Randomly draw $(g_t, i_j)$ from $\mathbf{Y}$ and sample negative examples for $g_t$ to constitute $\mathcal{D}_{g_t}$, and compute the group prediction loss $\mathcal{L}_{group}$ \textit{w.r.t.} Equation \ref{eq:group_loss};

      Randomly draw $(u_s, i_j)$ from $\mathbf{R}$ and sample negative examples for $u_s$ to constitute $\mathcal{D}_{u_s}$, and compute the user prediction loss $\mathcal{L}_{user}$ \textit{w.r.t.} Equation \ref{eq:user_loss};

      Take a gradient step to update $\Theta$ \textit{w.r.t.} $\mathcal{L}_{group} + \mathcal{L}_{user}$;
  }
\end{algorithm}

\section{Reproducibility}

\subsection{Datasets}
\subsubsection{Pre-processing}
Following previous works \cite{AGREE, SoAGREE, HCR}, we filter out the groups which have at least 2 members and have interacted with at least 3 items for pre-processing. Since both datasets only contain positive instances, \textit{i.e.}, observed interactions, we randomly sample from missing data as negative instances to pair with each positive instance.

\subsubsection{Crawling Process\label{sec:Crawl}} For the group page in Mafengwo, the group's traveling locations as well as joined members' information can be directly extracted. Then each member's personal visiting records can be obtained from his/her personal page. In this way, we can collect the group-level interacted locations and each member's visited locations. Repeatedly, a new dataset can be constructed that contains rich semantic information of items. Finally, we get a new dataset consisting of 11,027 users, 1,215 groups, and 1,236 items. This dataset is named ``Mafengwo-S'' with ``S'' referring to semantics.

\subsection{Implementation\label{sec:implement}}
 We implement ConsRec in PyTorch and optimize with Adam optimizer. For the initialization of the embedding layer, we apply the Glorot initialization strategy \cite{Glorot}. For hidden layers, we randomly initialize  their  parameters  with  a  Gaussian  distribution  of  a mean of 0 and a standard deviation of 0.1. For hyperparameters, we tune the number of convolutional layers $L$ in $\{ 1, 2, 3, 4 \}$ and the number of negative instances in $\{2,4,6,8,10,12 \}$. We empirically set the embedding dimension of 32. We implement the $\textbf{MLP}(\cdot)$ in prediction layer with a 3-layer setting and ReLU activation.

\subsection{Baselines\label{sec:baseline}}

We compare ConsRec with the following baselines: 
\begin{itemize}[leftmargin=*, topsep=0pt]
  \item \textbf{Popularity} \cite{Popularity} is a non-personalized method to benchmark the performance of other personalized methods. It recommends items to users and groups based on the popularity of items, which is measured by the number of interactions in the training set.
  \item \textbf{NCF} \cite{NCF} is adopted as treating each group as a virtual user for prediction. 
  \item \textbf{AGREE} \cite{AGREE} is a classical attentive aggregation-based group recommendation solution that selectively aggregates user representations within a group. 
  
    \item \textbf{HyperGroup} \cite{HyperGroup} models the group as hyperedge and proposes a hyperedge embedding-based representation learning method.
    
  \item \textbf{HCR} \cite{HCR} is a strong group recommendation baseline that proposes a dual channel hypergraph convolutional network to capture member-level and group-level preferences.

  \item \textbf{GroupIM} \cite{GroupIM} aggregates users’ preferences as group preferences via the attention mechanism. Particularly, to alleviate the data sparsity issue, it adds an extra self-supervised learning objective by maximizing the mutual information between the users and their belonging groups.

   \item \textbf{$\mathbf{S}^2$-HHGR} \cite{HHGR} is another strong baseline that integrates hypergraph learning and self-supervised learning for better group preferences' prediction.
  \item \textbf{CubeRec} \cite{HyperCube} is the state-of-the-art deep model in group recommendation. It utilizes the geometric expressiveness of hypercubes to adaptively aggregate members' interests. 
\end{itemize}

 For conducting the performance comparison, we use their official codes released at Github: NCF\footnote{https://github.com/hexiangnan/neural\_collaborative\_filtering}, AGREE\footnote{https://github.com/LianHaiMiao/Attentive-Group-Recommendation}, HCR\footnote{https://github.com/GroupRec/GroupRec}, GroupIM\footnote{https://github.com/CrowdDynamicsLab/GroupIM}, $\mathbf{S}^2$-HHGR\footnote{https://github.com/0411tony/HHGR}, and CubeRec\footnote{https://github.com/jinglong0407/CubeRec}. Since the source codes of HyperGroup \cite{HyperGroup} are unavailable, we implement this method based on its original paper. Actually, the released codes of some baselines are either problematic or incomplete, so we refactor their original codes.

To boost the development of group recommendation, we release ConsRec at \textcolor{blue}{https://github.com/FDUDSDE/WWW2023ConsRec}. We also release our implementations of above group recommendation baselines. For all baseline models, we refer to their best parameter set-ups reported in the original papers. If we use the same datasets and evaluation settings, we directly report their results. We conduct all the experiments on GPU machines of Nvidia Tesla V100 with 32GB memory.

\section{Experiments}
\begin{figure}[!t]
    \centering   
      \includegraphics[width=6cm]{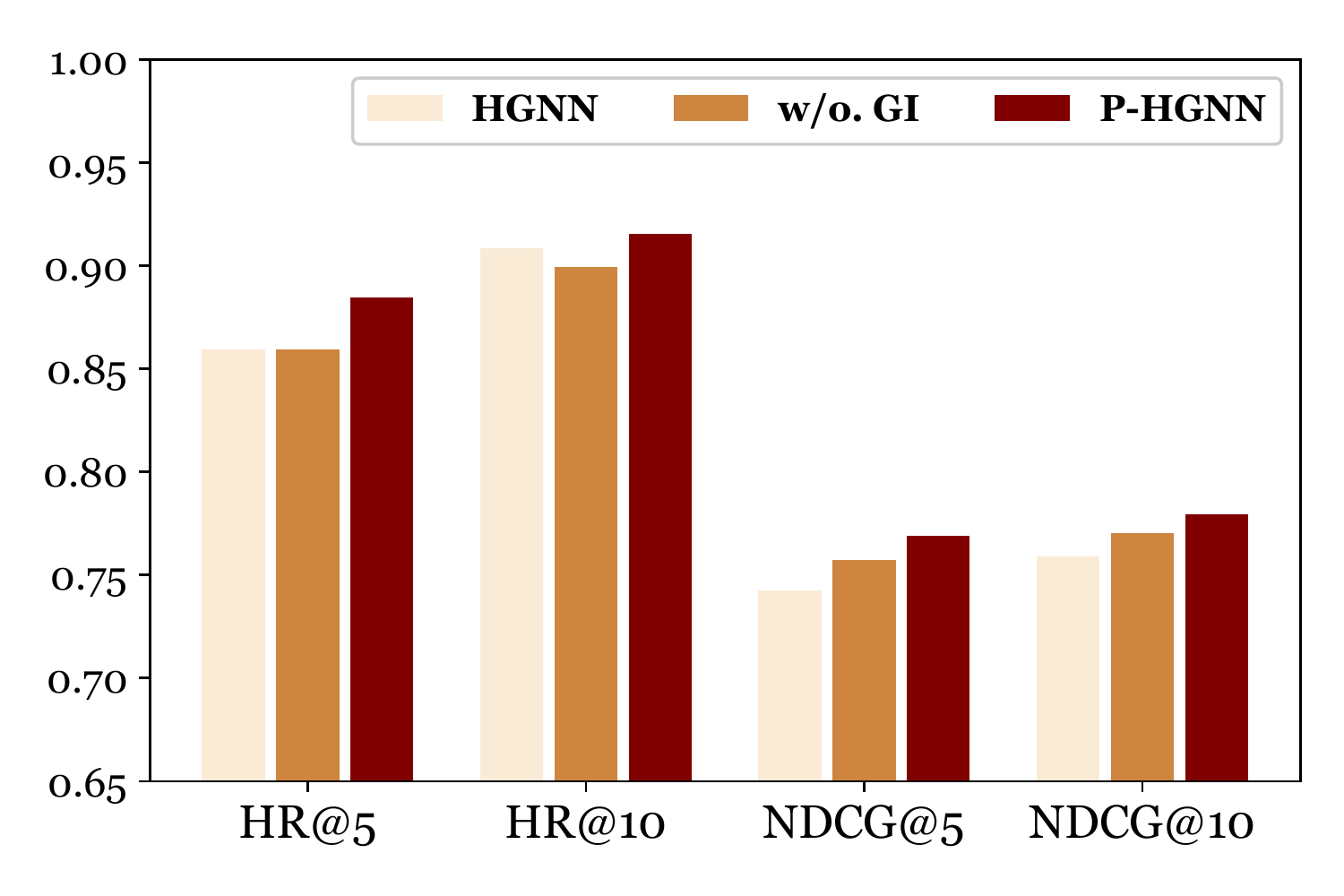}  
\caption{Ablation study on different hypergraph neural networks on Mafengwo with group recommendation results reported. }  
\label{fig:hgnn}   
\end{figure}

\begin{figure}[!t]
   \centering  
   \includegraphics[width=8cm]{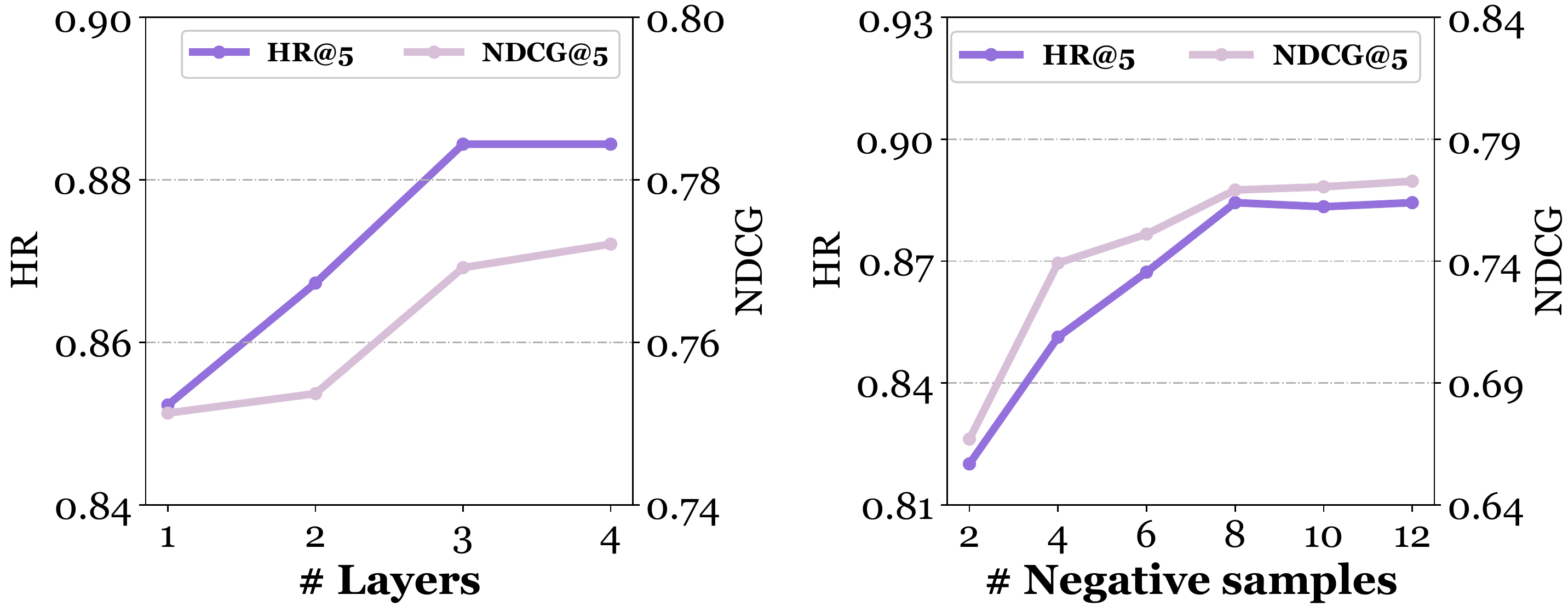}  
\caption{Parameters study on group recommendation task on Mafengwo.}   
\label{fig:param}   
\end{figure}

\subsection{Effectiveness of P-HGNN (RQ4)}
 Recall that we propose a preference-aware hypergraph neural network to yield a more expressive member-level aggregation. The concrete computation mechanism is shown in Figure \ref{fig:aggregation_final} and Equations \ref{eq: fusion} and \ref{eq:item_update}. To verify its effectiveness, we conduct the ablation study by comparing it with two different variants. One is the basic hypergraph neural network (HGNN) \cite{HyperGCN}. The other variant eliminates group-item element-wise product as $\mathbf{m}_e = \textbf{CONCAT}(\mathbf{m}_{e, u},  \mathbf{m}_{e,i}) \mathbf{W}^f$. This variant is denoted as ``w/o. GI''. 

 We show the experimental results in Figure \ref{fig:hgnn}. From this figure, ``w/o. GI'' variant realizes performance improvement compared with ``HGNN'', showing the necessity of preserving the distinct semantics of users and items. Our P-HGNN outperforms ``w/o. GI'', indicating the effectiveness of the introduction of group-item element-wise products. In a word, P-HGNN has the advantage of generating a more meaningful member-level aggregation, thus reinforcing groups' representations.

\subsection{Parameters Study (RQ5)}
In this subsection, we investigate the influence of two key parameters in our model, \textit{i.e.}, the number of (hyper)graph convolutional layers $L$ and the number of negative samples. 
\subsubsection{Number of Convolutional Layers}
The performance of graph convolutional network is affected by the number of graph convolution layers. As the number increases, the convolutional networks are faced with the problem of over-smoothing \cite{oversmooth} where node representations are not discriminative enough. To illustrate its influence, we show the performance \textit{w.r.t.} the number of convolutional layers in Figure \ref{fig:param}. We observe that when the layer is 3, better results can be obtained on Mafengwo. Therefore, we choose 3 as a default setting.

\subsubsection{Number of Negative Samples}
The strategy of negative sampling has been proven rational and effective in \cite{NegativesEffect}. It randomly samples various numbers of missing data as negative samples to pair with each positive instance. To illustrate the impact of negative sampling for our model, we show the performance of ConsRec \textit{w.r.t.} different numbers of negative samples in Figure \ref{fig:param}. We can observe that too small negative samples are not enough for optimization. With the increase of negative samples, the model performance firstly  improves and then becomes stable. Therefore, we choose 8 as a default setting.

\section{Discussions About Consensus}

Though consensus has also been studied in group decision making (GDM) task \cite{GDM1, GDM2}, here we point out the differences between GDM task and group recommendation (GR) task to show why GDM methods can not be directly applied for consensus modeling in GR scenarios.

\begin{itemize}[leftmargin=*, topsep=5pt]
    \item Task setting: GDM necessitates each member's preferences on all alternatives (\textit{i.e.}, items) as input, which relies on manual labeling by domain expertise. Instead, GR task automatically learns user-level preferences from user historical behaviors.
    \item Application scenario: Due to the heavy reliance on human efforts, GDM only works for some specific groups to choose from limited choices. Differently, GR task can deal with large-scale datasets that contain huge amounts of items and groups.
\end{itemize}

\end{document}